\documentclass[twocolumn,showpacs,amsmath,amssymb]{revtex4}

\usepackage{graphicx}
\usepackage{dcolumn}
\usepackage{bm}
\usepackage{amssymb}
\usepackage{amsmath}

\usepackage{epsf}
\usepackage{subfigure}
\usepackage{epstopdf}
\newcommand{\bra}[1]{\langle #1|}
\newcommand{\ket}[1]{|#1\rangle}
\newcommand{\braket}[2]{\langle #1|#2\rangle}

\newcommand{\mean}[1]{\langle #1 \rangle}
\newcommand{\trace}{{\rm Tr}}

\newcommand{\sbra}[1]{\ll #1|}
\newcommand{\sket}[1]{|#1 \gg}
\newcommand{\sbraket}[2]{\ll #1|#2 \gg}
\renewcommand{\i}{{\rm i}}
\newcommand{\e}{{\rm e}}
\begin{document}

\title{Fluctuation theorems for quantum master equations}

\author{Massimiliano Esposito} 
\altaffiliation[Also at ]{Center for Nonlinear Phenomena and Complex Systems,\\
Universit{\'e} Libre de Bruxelles, Code Postal 231, Campus Plaine, 
B-1050 Brussels, Belgium.}
\author{Shaul Mukamel}
\affiliation{Department of Chemistry, University of California, 
Irvine, California 92697, USA.}
\date{\today}

\begin{abstract}
A quantum fluctuation theorem for a driven quantum subsystem
interacting with its environment is derived based solely on 
the assumption that its reduced density matrix obeys a closed 
evolution equation i.e. a quantum master equation (QME).
Quantum trajectories and their associated entropy, heat and work 
appear naturally by transforming the QME to a time dependent 
Liouville space basis that diagonalizes the instantaneous
reduced density matrix of the subsystem. 
A quantum integral fluctuation theorem, a steady state fluctuation 
theorem and the Jarzynski relation are derived in a similar way as 
for classical stochastic dynamics.
\end{abstract}



\maketitle
\section{Introduction \label{intro}}

The fluctuation theorems and the Jarzynski relation are some of a handful
of powerful results of nonequilibrium statistical mechanics that hold 
far from thermodynamic equilibrium.
Originally derived in the context of classical mechanics 
\cite{Jarzynski1}, the Jarzynski relation has been subsequently 
extended to stochastic dynamics \cite{Crooks0}.
It relates the distribution of the work done by a driving force 
of arbitrary speed on a system initially at equilibrium 
(nonequilibrium property) to the free energy difference between the 
initial and final equilibrium state of the system (equilibrium property). 
This remarkable relation has recently been shown to hold for arbitrarily 
coupling strength between the system and the environment (see Jarzynski's 
reply \cite{JarzynskiReply} to criticism from Ref. \cite{Cohen}). 
The fluctuation theorems are based on a fundamental relation connecting 
the entropy production of a single system trajectory to the logarithm
of the ratio of the probability of the forward and the backward 
trajectory \cite{Maes3}. 
The ensemble average of the trajectory entropy production is the 
macroscopic entropy production of the system whereas its distribution
gives rise to various kinds of fluctuation theorems. 
The first has been derived for classical mechanics and initially 
for deterministic (but non-Hamiltonian) 
thermostated systems \cite{Evans1,Gallavotti,Evans2}. 
Some interesting studies of fluctuation relations valid for far from 
equilibrium classical Hamiltonian systems have been made
even earlier \cite{Bochkov,Stratonovich,JarsynskiFT}.
Fluctuation theorems for systems with stochastic dynamics have also 
been developed 
\cite{Kurchan1,Lebowitz,Searles,Gaspard1,Gaspard2,Crooks1,Crooks2,Seifert}.
For classical stochastic dynamics, the connection between the fluctuation 
theorem and the Jarzynski relation has been established by Crooks 
\cite{Crooks1}.
Seifert has recently provided a unified description of the different
fluctuation relations and of the Jarzynski relation for classical
stochastic processes described by master equations \cite{Seifert}. \\
The understanding of these two fundamental relations in quantum 
mechanics is still not fully established. 
Quantum Jarzynski relations have been investigated in 
\cite{HTasaki,Yukawa,Mukamel,Maes}.
Quantum fluctuation theorems have been developed only in a few 
restricted situations \cite{Kurchan,STasaki,Maes1,Monnai}. 
A quantum exchange fluctuation theorem has also been considered in 
\cite{JarzynskiQ}.
Some interesting considerations on the quantum definition of work in the 
previous studies have been made in \cite{Allahverdyan}. \\
It should be noted that the dynamics of an isolated (whether driven or not) 
quantum system is unitary and its von Neumann entropy is time independent. 
Therefore, fluctuation theorems for such closed systems are useful 
only provided one defines some reduced macrovariable dynamics or some 
measurement process on the system \cite{Maes2}.\\ 

The purpose of this paper is to provide a unified derivation for the
different quantum fluctuation relations (an integral fluctuation theorem, 
a steady state fluctuation theorem and the Jarzynski relation). 
We build upon the unification of the different fluctuation relations 
recently accomplished by Seifert \cite{Seifert} for classical stochastic 
dynamics described by birth and death master equation (BDME).  
Quantum evolution involves coherences which make its interpretation in
term of trajectories not obvious.
Nevertheless, we show that it is possible to formally develop a 
trajectory picture of quantum dynamics which allows to 
uniquely represent entropy, heat and work distributions.
This rely on the single assumption that the reduced 
dynamics of a driven quantum subsystem interacting with its 
environment is described by a closed evolution equation for 
the density matrix of the subsystem i.e. a QME 
\cite{Haake,Spohn,Gardiner,Breuer}. 
However, while the physical quantities defined along classical trajectories 
are conceptually clear and experimentally measurable, how to measure 
the physical quantities associated to quantum trajectories remains a 
fascinating open issue intimately connected to quantum measurement.\\

The plan of the paper is as follows: 
We start in section \ref{heatwork1} by defining quantum heat 
and quantum work for a driven subsystem interacting with its 
environment in consistency with thermodynamics .
We then discuss the consequences of defining heat and work in 
terms of the time dependent basis which diagonalizes the 
subsystem density matrix in section \ref{heatwork2}. 
In section \ref{DBMEformulation}, we show that by assuming a QME 
for the subsystem reduced density matrix we can recast 
its solution in a representation which takes the form of a BDME 
with time dependent rates. 
In section \ref{entropy}, we show that the BDME representation allows 
to split the entropy evolution in two parts, the entropy flow associated 
with exchange processes with the environment and the entropy production 
associated with subsystem internal irreversible processes.
In section \ref{trajectory}, we show that the BDME representation 
naturally allows to define quantum trajectories as well as 
their associated entropy flow and entropy production. 
We then derive the fundamental relation of this paper (\ref{1Aaabgi}) 
which will allow us to derive, in section \ref{integral}, a quantum integral 
fluctuation theorem and, in section \ref{steady}, a quantum steady 
state fluctuation theorem. 
Having identified in section \ref{trajQW} the heat and the work 
associated to the quantum trajectories, we show in section \ref{Jarzynski}
that the fundamental relation of section \ref{trajectory} also allows to 
derive a quantum Jarzynski relation.
We finally draw conclusions in section \ref{conclusion}.

\section{Average heat and work \label{heatwork1}}

We start by defining the average quantum heat and work for a driven 
subsystem interacting with its environment and show the consistency 
of these definitions with thermodynamics . 
Heat and work can be rigorously expressed 
in term of the reduced density matrix of the subsystem without having 
to refer explicitly to the environment.\\

We consider a driven subsystem with Hamiltonian $\hat{H}_S(t)$. 
Everywhere in this paper we denote operators with an hat (and superators 
with two hats) and we use the Schrodinger picture where the time 
dependence of the observables is explicit and comes exclusively from 
external driving.
We could also have written $\hat{H}_S(\lambda(t))$, where $\lambda(t)$ 
is the external time dependent driving.
This subsystem is interacting with its environment whose Hamiltonian is 
$\hat{H}_B$. The interaction energy between the subsystem and the environment
is described by $\hat{H}_I$.
The Hamiltonian of the total system reads therefore 
\begin{eqnarray}
\hat{H}_{T}(t)=\hat{H}_S(t)+\hat{H}_B+\hat{H}_I \; .
\label{2Aaaaa}
\end{eqnarray}
We have assumed that the driving acts exclusively on the subsystem 
and does not affect $\hat{H}_B$ and $\hat{H}_I$.\\
The state of the total system is described by the density matrix 
$\hat{\rho}(t)$ which obeys the von Neumann equation
\begin{eqnarray}
\dot{\hat{\rho}}(t)=-i[\hat{H}_{T}(t),\hat{\rho}(t)]
=\hat{\hat{{\cal L}}}(t) \hat{\rho}(t) \;,
\label{2Aaaaab}
\end{eqnarray}
The energy of the total system is given by
\begin{eqnarray}
\mean{\hat{H}_{T}}_t \equiv \trace \hat{H}_T(t) \hat{\rho}(t)  \; .
\label{2Aaaab}
\end{eqnarray}
The change in the total energy between time $0$ and $t$ due 
to the time dependent driving is therefore given by
\begin{eqnarray}
\Delta E_T(t) \equiv 
\int_0^t d\tau \frac{d \mean{\hat{H}_{T}}_{\tau}}{d\tau}  
= W_T(t) + Q_T(t) \; ,
\label{2Aaaac}
\end{eqnarray}
where the work and the heat have respectively been defined as
\begin{eqnarray}
W_T(t) \equiv \int_0^t d\tau \trace \dot{\hat{H}}_{T}(t) \hat{\rho}(t) 
\label{2Aaaada}\\
Q_T(t) \equiv \int_0^t d\tau \trace \hat{H}_{T}(t) \dot{\hat{\rho}}(t) \; .
\label{2Aaaadaa}
\end{eqnarray}
Using the von Neumann equation (\ref{2Aaaaab}) and the invariance 
of the trace under cyclic permutation \cite{Comment}, we find that 
no heat is generated in the isolated total system  
\begin{eqnarray}
Q_T(t) = -i \int_0^t d\tau\trace \hat{H}_{T}(t) [\hat{H}_{T}(t),\hat{\rho}(t)] = 0 \;.
\label{2Aaaadb}
\end{eqnarray}
We next turn to the subsystem. 
Its reduced density matrix is defined as 
$\hat{\sigma}(t) \equiv \trace_B \hat{\rho}(t)$ and its energy is given by
\begin{eqnarray}
\mean{\hat{H}_S}_t \equiv \trace \hat{H}_S(t) \hat{\rho}(t) 
= \trace_S \hat{H}_S(t) \hat{\sigma}(t) \;.
\label{2Aaaaeb}
\end{eqnarray}
The change in this energy between time $0$ and $t$ is given by
\begin{eqnarray}
\Delta E_S(t) \equiv \int_0^t d\tau \frac{d \mean{\hat{H}_{S}}_{\tau}}{d\tau} 
= W_S(t) + Q_S(t) \; ,
\label{2Aaaaf}
\end{eqnarray}
where the work and the heat are defined as
\begin{eqnarray}
W_S(t) &\equiv& \int_0^t d\tau \trace \dot{\hat{H}}_S(\tau) \hat{\rho}_T(\tau) \label{2Aaaaga}\\
&=& \int_0^t d\tau \trace_S \dot{\hat{H}}_S(\tau) \hat{\sigma}(\tau) \nonumber \\
Q_S(t) &\equiv& \int_0^t d\tau \trace \hat{H}_S(\tau) \dot{\hat{\rho}}_T(\tau) \label{2Aaaagaa}\\
&=& \int_0^t d\tau \trace_S \hat{H}_S(\tau) \dot{\hat{\sigma}}(\tau) \nonumber \;.
\end{eqnarray}
Since the time dependence of the total system Hamiltonian comes solely from the 
subsystem Hamiltonian, $\dot{\hat{H}}_B=\dot{\hat{H}}_I=0$, $\dot{\hat{H}}_T=\dot{\hat{H}}_S$ and 
the work done by the driving force on the subsystem is the same as the work 
done by this force on the total system 
\begin{eqnarray}
W_T(t) = W_S(t) \equiv W(t) \;.
\label{2Aaaagb}
\end{eqnarray}
This also means that the energy increase in the subsystem minus the amount 
of heat which went to the environment is equal to the energy increase
in the total system 
\begin{eqnarray}
W(t) = \Delta E_T(t) = \Delta E_S(t) - Q_S(t) \; .
\label{2Aaaah}
\end{eqnarray}
It should be noticed that due to the absence of heat flux in the total system 
$Q_T(t)=0$, using (\ref{2Aaaadaa}) with (\ref{2Aaaaa}) and (\ref{2Aaaagaa}), 
we can also express the heat going from the subsystem to the environment as
\begin{eqnarray}
Q_S(t) = -\int_0^t d\tau \frac{d \mean{\hat{H}_{B}}_{\tau}}{d\tau} 
- \int_0^t d\tau \frac{d \mean{\hat{H}_{I}}_{\tau}}{d\tau} \; .
\label{2Aaaai}
\end{eqnarray}

\section{Calculating heat and work in a time dependent basis \label{heatwork2}}

As will become clear in section \ref{DBMEformulation}, in order to 
associate trajectories to the quantum dynamics, one need to represent 
the dynamics in a time dependent basis.
This is a fundamental difference from classical thermodynamics where
the basis set (coordinate system) is fixed. 
In order to associate heat and work with single trajectories 
they must be defined with respect to the time dependent basis set.
The ensemble average of the quantities defined for the trajectories
will therefore also depend on the time dependent basis set.
For this reason, we introduce a modified definition of heat and work.
The effect of the basis time dependence on heat and work is given in 
appendix \ref{appQW}.\\

The energy of the total system (\ref{2Aaaab}) can also be written as
\begin{eqnarray}
\mean{\hat{H}_{T}}_t = \sum_{\alpha}  P^T_t(\alpha)
\bra{\alpha_t} \hat{H}_T(t) \ket{\alpha_t}\; ,
\label{1Aaaab}
\end{eqnarray}
where we have introduced the time dependent basis $\{\ket{\alpha_t}\}$
which diagonalizes the instantaneous density matrix at all time
\begin{eqnarray}
\bra{\alpha_t} \hat{\rho}(t) \ket{\alpha_{t}'} =
\bra{\alpha_t} \hat{\rho}(t) \ket{\alpha_{t}} \delta_{\alpha \alpha'} 
\equiv P^T_t(\alpha) \delta_{\alpha \alpha'} \;.
\label{1Aaaabb}
\end{eqnarray}
The change in this energy between time $0$ and $t$ due to the time 
dependent driving can therefore be rewritten as
\begin{eqnarray}
\Delta E_T(t) = \tilde{W}_T(t) + \tilde{Q}_T(t) \; ,
\label{1Aaaac}
\end{eqnarray}
where the modified work and heat have respectively 
been defined as
\begin{eqnarray}
\tilde{W}_T(t) &\equiv& \int_0^t d\tau \sum_{\alpha} P^T_{\tau}(\alpha)
\frac{d}{d\tau} \bigg( \bra{\alpha_{\tau}} \hat{H}_T(\tau) \ket{\alpha_{\tau}} 
\bigg) \label{1Aaaada}\\
\tilde{Q}_T(t) &\equiv& \int_0^t d\tau \sum_{\alpha} \dot{P}^T_{\tau}(\alpha)
\bra{\alpha_{\tau}} \hat{H}_T(\tau) \ket{\alpha_{\tau}} \;.
\label{1Aaaadaa}
\end{eqnarray}
Because the total system is driven but otherwise isolated, its 
evolution is unitary and we have (see appendix \ref{appQW}) 
\begin{eqnarray}
\tilde{W}_T(t) &=& W_T(t) = \Delta E_T(t) \label{1Aaaadax}\\
\tilde{Q}_T(t) &=& Q_T(t) = 0 \;. \label{1Aaaadaax}
\end{eqnarray}
This means that defining heat and work on the time dependent basis which 
diagonalizes the instantaneous density matrix for unitary evolution is 
equivalent to the original definition of heat and work in a time 
independent basis.\\

The energy of the subsystem (\ref{2Aaaaf}) can also be 
written in analogy to (\ref{1Aaaab}) as
\begin{eqnarray}
\mean{\hat{H}_S}_t = \sum_{m} P_t(m) \bra{m_t} \hat{H}_S(t) \ket{m_t} , 
\label{1Aaaaeb} 
\end{eqnarray}
where we have introduced the time dependent basis $\{ \ket{m_t} \}$
diagonalizing the instantaneous subsystem reduced density matrix 
\begin{eqnarray}
\bra{m_t} \hat{\sigma}(t) \ket{m_{t}'} =
\bra{m_t} \hat{\sigma}(t) \ket{m_{t}} \delta_{m m'} 
\equiv P_t(m) \delta_{m m'} \;.\label{1Aaaaec}
\end{eqnarray}
Let us note for future reference that
\begin{eqnarray}
\frac{d}{dt} ( \bra{m_t} \hat{\sigma}(t) \ket{m_t'} ) &=& 
\bra{m_t} \dot{\hat{\sigma}}(t) \ket{m_t'} \label{1Aaaagaaa}\\
&&\hspace{-0cm}+\bra{\dot{m}_t}  \hat{\sigma}(t) \ket{m_t'} 
+ \bra{m_t} \hat{\sigma}(t) \ket{\dot{m}_t'} \;. \nonumber 
\end{eqnarray}
Eq. (\ref{1Aaaaec}) and (\ref{1Aaaagaaa}) give 
\begin{eqnarray}
\bra{m_t} \dot{\hat{\sigma}}(t) \ket{m_t'} &=& 
\dot{P}_t(m) \delta_{mm'} \label{1Aaaagaaab}\\
&&\hspace{-0cm}- \braket{\dot{m}_t}{m_t'} P_t(m')
- \braket{m_t}{\dot{m}_t'} P_t(m) \;. \nonumber
\end{eqnarray}
Notice also that for $m=m'$, we have
\begin{eqnarray}
\bra{m_t} \dot{\hat{\sigma}}(t) \ket{m_t}
=\dot{P}_t(m) \label{1Aaaagaaac}
\end{eqnarray}
because $\braket{\dot{m}_t}{m_t} + \braket{m_t}{\dot{m}_t}
= \frac{d}{dt} ( \braket{m_t}{m_t} ) =0$.\\
Using (\ref{1Aaaaeb}), the change in the subsystem energy between 
time $0$ and $t$ can be rewritten as 
\begin{eqnarray}
\Delta E_S(t) = \tilde{W}_S(t) + \tilde{Q}_S(t) \; , \label{1Aaaaf}
\end{eqnarray}
where the work and the heat are defined in analogy 
to (\ref{1Aaaada}) and (\ref{1Aaaadaa}) as
\begin{eqnarray}
\tilde{W}_S(t) &\equiv& \int_0^t d\tau \sum_{m} P_{\tau}(m) 
\frac{d}{d\tau} \bigg( \bra{m_{\tau}} \hat{H}_S(\tau) \ket{m_{\tau}} \bigg)
\label{1Aaaaga} \\
\tilde{Q}_S(t) &\equiv& \int_0^t d\tau \sum_{m} 
\dot{P}_{\tau}(m) \bra{m_{\tau}} \hat{H}_S(\tau) \ket{m_{\tau}}
\label{1Aaaagaa} \;.
\end{eqnarray}
It is shown in appendix \ref{appQW} that the work and heat
defined in the time dependent basis $\{ \ket{m_t} \}$ is related
to the original work and heat defined in any time independent 
basis by
\begin{eqnarray}
\tilde{W}_S(t) &=& W_S(t) + A_S(t) \label{1Aaaagaac} \\
\tilde{Q}_S(t) &=& Q_S(t) - A_S(t) \label{1Aaaagaad} \;,
\end{eqnarray}
where 
\begin{eqnarray}
A_S(t) &\equiv& 
\int_0^t d\tau \sum_{m} P_{\tau}(m) \label{1Aaaagaab} \\
&&\hspace{1cm} \bigg( \bra{\dot{m}_{\tau}} \hat{H}_S(\tau) \ket{m_{\tau}} 
+ \bra{m_{\tau}} \hat{H}_S(\tau) \ket{\dot{m}_{\tau}} \bigg) \; .\nonumber
\end{eqnarray}
It should be emphasized that both the original and the modified work 
and heat of the subsystem can be defined exclusively in term of the 
subsystem quantities without refering explicitly to the environment.\\   
Using (\ref{2Aaaagb}) and (\ref{2Aaaah}) with (\ref{1Aaaagaac}) and 
(\ref{1Aaaagaad}), we get
\begin{eqnarray}
\Delta E_T(t) &=& W(t) = \tilde{W}_S(t) - A_S(t) \label{1Aaaagb}\\
&=& \Delta E_S(t) - \tilde{Q}_S(t) - A_S(t) \nonumber \;.
\end{eqnarray}

\section{BDME representation of the QME solution \label{DBMEformulation}}

In this section we show that if we assume a closed evolution equation 
for the subsystem reduced density matrix, we can transform it solution
in a BDME form with time dependent rates.\\

We assume that the reduced subsystem density matrix 
$\hat{\sigma}(t)$ obeys a closed QME. The literature on this topic is 
well furnished \cite{Haake,Spohn,Gardiner,Breuer}. 
This QME can be derived microscopically by perturbation theory 
like in the Redfield theory or using a quantum dynamical 
semigroups approach leading to Lindblad type master equations. 
In Liouville space \cite{MukamelB}, the QME of the externally driven 
subsystem interacting with its environment reads 
\begin{eqnarray}
\sket{\dot{\hat{\sigma}}(t)}= 
\hat{\hat{{\cal K}}}(t) \sket{\hat{\sigma}(t)} \;.
\label{1Aaaaja}
\end{eqnarray}
If the interaction with the environment vanishes, the generator 
$\hat{\hat{{\cal K}}}(t)$ becomes the antihermitian superoperator 
$\hat{\hat{{\cal K}}}(t)=\hat{\hat{{\cal L}}}_S(t)=
-\i \lbrack \hat{H}_S(t),\cdot \rbrack$ 
and the evolution superoperator $\hat{\hat{{\cal M}}}_t$ defined by 
$\sket{\hat{\sigma}(t)}=\hat{\hat{{\cal M}}}_t \sket{\hat{\sigma}(0)}$ 
becomes the unitary superoperator 
$\hat{\hat{{\cal M}}}_t=\exp_{+}{\{\int_{0}^{t} d\tau 
\hat{\hat{{\cal L}}}_S(\tau)\}}$. 
However, for non vanishing coupling this generator is not 
antihermician and leads to a nonunitary evolution. \\
The QME in some given (possibly time dependent) basis reads
\begin{eqnarray}
\sbraket{ii'}{\dot{\hat{\sigma}}(t)}
= \sum_{jj'} \sbra{ii'}\hat{\hat{{\cal K}}}(t)\sket{jj'} 
\sbraket{jj'}{\hat{\sigma}(t)} \;,
\label{1Aaaajb}
\end{eqnarray}
where $\sbraket{jj'}{\hat{\sigma}(t)}$ is the superoperator 
representation of $\bra{j} \hat{\sigma}(t) \ket{j'}$.
Let us now use the time dependent basis $\{\ket{m_t}\}$ 
introduced in (\ref{1Aaaaec}).
Since the QME keeps $\hat{\sigma}(t)$ hermitian, this diagonalization 
is always possible.
\begin{eqnarray}
\sbraket{m_t m_t'}{\hat{\sigma}(t)} = P_t(m)\delta_{m,m'} \;.
\label{1Aaaak}
\end{eqnarray}
A crucial property of this basis is that [see Eq. (\ref{1Aaaagaaac})]
\begin{eqnarray}
\dot{P}_t(m) 
= \sbraket{m_t m_t}{\dot{\hat{\sigma}}(t)} \;.
\label{1Aaaal}
\end{eqnarray}
The consequence of this property is that by defining
\begin{eqnarray}
W_t(m',m) &\equiv& \sbra{m_t m_t} \hat{\hat{{\cal K}}}(t) \sket{m_t' m_t'} \;,
\label{1Aaaan}
\end{eqnarray}
and by projecting the QME (\ref{1Aaaaja}) on the time dependent 
superbra $\sbra{m(t)m(t)}$ we get  
\begin{eqnarray}
\dot{P}_t(m) = \sum_{m'} W_t(m',m) P_t(m') \;.
\label{1Aaaao}
\end{eqnarray}
Since the QME (\ref{1Aaaaja}) preserves probability, 
we have $\sum_{m} W_t(m',m)=0$ and $W_t(m',m)$ real.
Therefore, we can rewrite (\ref{1Aaaao}) as 
\begin{eqnarray}
\dot{P}_t(m) = \sum_{m'\neq m} \{ W_t(m',m) P_t(m') 
- W_t(m,m') P_t(m) \} \;.
\label{1Aaaap}
\end{eqnarray}
This equation appears like a BDME but should not be viewed 
as an equation of motion. 
It is merely a way of recasting the solution of the QME 
(\ref{1Aaaaja}) in a diagonal basis.
In fact, in order to get the $P_t(m)$'s and the $W_t(m',m)$'s, 
we need to solve the QME first and find the time dependent 
unitary transformation diagonalizing the solution $\hat{\sigma}(t)$ 
at any time. 
Eq. (\ref{1Aaaao}) should therefore be viewed as a formal 
definition of the rate matrix $W_t(m',m)$.
We will show that $W_t(m',m)$ defined in this way can be used 
to derive quantum fluctuation relations.
Note that $W_t(m',m)$ depends on the subsystem initial 
condition $\hat{\sigma}(0)$.\\

If the subsystem (driven or not) does not 
interact with the environment, the generator is antihermician 
and the evolution superoperator unitary. In this case 
$\sket{\hat{\sigma}(t)} = \hat{\hat{{\cal M}}}_t \; \sket{\hat{\sigma}(0)}$
and $\sbra{m_t m_t} = \sbra{m_0 m_0} \; \hat{\hat{{\cal M}}}^{-1}_t$, so that
\begin{eqnarray}
P_t(m) = \bra{m_0} \hat{\sigma}(0) \ket{m_0} = P_0(m) \;.
\label{1Aaaapb}
\end{eqnarray}
This shows that the $P_t(m)$'s evolve only if the dynamics is 
nonunitary.\\ 
When there is no driving and the subsystem does interact with its 
environment, the dynamics is nonunitary and the subsystem will 
reach equilibrium $\hat{\sigma}^{\rm eq}$ on long time scales. 
For an infinite isothermal environment this equilibrium state will
correspond to the canonical subsystem reduced density matrix 
$\hat{\sigma}^{\rm eq}=\e^{-\beta \hat{H}_S}/Z_S$ where 
$Z_S=\trace \; \e^{-\beta \hat{H}_S}$ and $\beta=1/T$ ($k_B \equiv 1$). 
In this case the basis diagonalizing $\hat{\sigma}^{\rm eq}$ becomes time 
independent and will also diagonalize the subsystem Hamiltonian
so that $P^{\rm eq}(m)=e^{-\beta E_m}/Z_S$ where $E_m$ are the 
eigenvalues of the subsystem Hamiltonian.\\
For a subsystem with nonequilibrium boundary conditions and 
interacting with its environment, the subsystem can reache a 
steady state $\hat{\sigma}^{\rm st}$ on long times. 
In this case the matrix diagonalizing the density matrix is again 
time independent and both the probabilities $P_t(m)=P^{\rm st}(m)$ 
and the rates $W_t(m',m)=W^{\rm st}(m',m)$ become time independent.
 
\section{Entropy for quantum ensembles \label{entropy}}

In this section we define the von Neumann entropy associated 
with the subsystem and separate its evolution into two parts: the entropy 
flow associated to the heat going from the subsystem to the environment 
and the entropy production associated to the internal (always positive) 
entropy growth of the subsystem.\\

The von Neumann entropy of the subsystem is defined by
\begin{eqnarray}
S(t) \equiv -\trace \; \hat{\sigma}(t) \ln \hat{\sigma}(t) 
= - \sum_m P_t(m) \ln P_t(m)
\label{1Aaaaq}
\end{eqnarray}
Using Eq. (\ref{1Aaaap}), we can write its time derivative as 
\begin{eqnarray}
\dot{S}(t)&=& -\sum_m \dot{P}_t(m) \ln P_t(m) \\
&=&- \sum_{m,m'} P_t(m) W_t(m,m') \ln \frac{P_t(m')}{P_t(m)}
\label{1Aaaar}
\end{eqnarray}
In analogy with \cite{Lebowitz,Gaspard1} for classical systems, 
this can be partitioned as
\begin{eqnarray}
\dot{S}(t) = \dot{S}_e(t) + \dot{S}_i(t) \;,
\label{1Aaaas}
\end{eqnarray}
where 
\begin{eqnarray}
\dot{S}_e(t) \equiv - \sum_{m,m'} P_t(m) W_t(m,m') 
\ln \frac{W_t(m,m')}{W_t(m',m)} \;,
\label{1Aaaat}
\end{eqnarray}
and where
\begin{eqnarray}
\dot{S}_i(t) \equiv \sum_{m,m'} P_t(m) W_t(m,m') 
\ln \frac{P_t(m) W_t(m,m')}{P_t(m') W_t(m',m)} \;.
\label{1Aaaau}
\end{eqnarray}
As a consequence of the inequality $(R_1-R_2) \ln (R_1/R_2) \geq 0$,
we notice that $\dot{S}_i(t) \geq 0$ is an always positive quantity. 
We will therefore identify it with the entropy production.
The remaining part of the entropy $\dot{S}_e(t)$ is thus associated 
with the entropy flow to the environment since in thermodynamics the 
entropy evolution is partitioned in the (reversible) entropy flow to 
the environment and the (irreversible) entropy production 
\cite{GrootMazur,Prigogine}.
To further rationalize this identification, let us assume that 
$W_t(m,m')$ satisfy the detailed balance condition 
\cite{Crooks1,Kampen,GrootMazur}.
For isothermal environments at temperature $T$, the detailed 
balance condition with respect to $\hat{H}_s(t)$, which means 
that the non-driven subsystem tends to thermal equilibrium at 
long time, reads
\begin{eqnarray}
\frac{W_t(m,m')}{W_t(m',m)}=
\e^{\beta \left(\bra{m_t} \hat{H}_S(t) \ket{m_t}
- \bra{m_t'} \hat{H}_S(t) \ket{m_t'} \right)} \;.
\label{1Aaaay}
\end{eqnarray}
Noticing that the heat (\ref{1Aaaagaa}) can be rewritten as
\begin{eqnarray}
\dot{\tilde{Q}}(t)&=& \sum_m \dot{P}_t(m) 
\bra{m_t} \hat{H}_S(t) \ket{m_t} \label{1Aaaax} \\
&=& - \sum_{m,m'} P_t(m) W_t(m,m') \nonumber \\
&&\hspace{1cm}\left( \bra{m_t} \hat{H}_S(t) \ket{m_t} - 
\bra{m_t'} \hat{H}_S(t) \ket{m_t'} \right) \nonumber \\
&=& - T \sum_{m,m'} P_t(m) W_t(m,m') \ln 
\frac{\e^{\beta \bra{m_t} \hat{H}_S(t) \ket{m_t}}}
{\e^{\beta \bra{m_t'} \hat{H}_S(t) \ket{m_t'}}} \nonumber 
\end{eqnarray}
and using (\ref{1Aaaat}), the immediate consequence of 
(\ref{1Aaaay}) is that the entropy flow is equal to the modified 
heat going from the subsystem to the environment divided by the 
environment temperature as expected from thermodynamics
\begin{eqnarray}
\dot{S}_{e}(t) = \frac{\dot{\tilde{Q}}_S}{T} \;.
\label{1Aaaaw}
\end{eqnarray}
This motivates our partition of the entropy (\ref{1Aaaas}) 
and the definition of the modified heat in section \ref{heatwork2}.\\

We can further show that the entropy flow is associated to 
reversible entropy variations.
In the thermodynamical sense, a reversible transformation is a 
one during which the entropy production is zero 
$\dot{S}_i(t) = 0$.
This property holds provided the following condition is satisfied 
[see (\ref{1Aaaau})]
\begin{eqnarray}
P_t(m) W_t(m,m') = P_t(m') W_t(m',m) \;.
\label{1Aaaav}
\end{eqnarray}
Using now Eq.(\ref{1Aaaay}), we find that for a reversible transformation 
the subsystem has to be at all time in the time dependent state 
\begin{eqnarray}
P_t(m) = \frac{\e^{-\beta \bra{m_t} \hat{H}_S(t) \ket{m_t}}}
{\sum_m \e^{-\beta \bra{m_t} \hat{H}_S(t) \ket{m_t}}} \;.
\label{1Aaaavb}
\end{eqnarray}
This state correspond to the instantaneous Gibbs state of the 
subsystem $\hat{\sigma}(t)=\e^{-\beta \hat{H}_S(t)}/Z_S$. 
In this case $\{ \ket{m}_t \}$ in Eq.(\ref{1Aaaav}) becomes the adiabatic 
basis (basis diagonalizing the subsystem Hamiltonian).
We thus show that for reversible transformations the probability 
distribution remains Gibbsian along the adiabatic levels.
Because $\dot{S}_i(t) = 0$, we also have $\dot{S}(t)=\dot{S}_{e}(t)$. 
Using (\ref{1Aaaay}), this means that for a reversible transformation
the change in the entropy of the subsystem result exclusively from the
heat flow to the environment $\dot{S}(t)=\dot{\tilde{Q}}_S/T$ 
in consistency with thermodynamics.\\
When there is no driving, Eq. (\ref{1Aaaav}) with Eq. (\ref{1Aaaay}) 
define equilibrium. 
At equilibrium we have $\dot{S}(t)=\dot{S}_{i}(t)=\dot{S}_{e}(t)=0$.\\

\section{Entropy for quantum trajectories \label{trajectory}}

In this section we introduce quantum trajectories and distributions. 
We will associate an entropy with these trajectories and identify the 
entropy flow and production of these trajectories whose ensemble averages 
recover the entropies discussed in section \ref{entropy}. 
This will allow us to derive a fundamental quantum relation similar to the 
classical relation obtained by Crooks \cite{Crooks1} and Seifert \cite{Seifert} 
connecting the ratio of the probability of a forward trajectory and the 
"backward" one with the trajectory entropy production. \\ 

\begin{figure}[h]
\centering
\rotatebox{0}{\scalebox{0.45}{\includegraphics{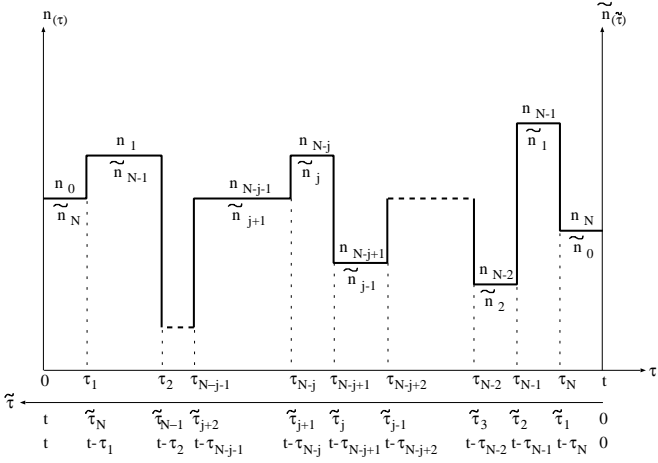}}}
\caption{Representation of a quantum forward trajectory $n_{(\tau)}$ 
and of the associated pseudo-backward trajectory $\tilde{n}_{(\tilde{\tau})}$.} 
\label{fig1}
\end{figure}

From Eq. (\ref{1Aaaap}) it seems natural to unravel the 
evolution equation for the probability $P_t(m)$ in the same way as
is done for classical stochastic processes \cite{Seifert}.
Let us consider a stochastic trajectory of duration $t$ which contains $N$ jumps. 
Different trajectories can of course have a different number of jumps $N$. 
$\tau=\lbrack 0,t \rbrack$ labels time during the process.
$j=1,\hdots,N$ labels the jumps.   
The trajectory $n_{(\tau)}$ [see Fig. \ref{fig1}] is made by the successive 
states taken by the system in time 
\begin{eqnarray}
n_{(\tau)} = n_0 \to n_1 \to n_2 \to \hdots \to n_N \;.
\label{1Aaaayb}
\end{eqnarray}
The system starts in $n_0$, jumps at time $\tau_j$ from 
$n_{j-1}$ to $n_{j}$ and ends up at time $t$ in $n_N$.
We will denote $\tau_0=0$ and $\tau_{N+1}=t$.\\
The entropy associated with the trajectory $n_{(\tau)}$ reads
\begin{eqnarray}
s(\tau) \equiv -\ln P_{\tau}(n_{(\tau)}) \;,
\label{1Aaaba}
\end{eqnarray}
where $P_{\tau}(n_{(\tau)})$ is the solution of Eq. (\ref{1Aaaap}) for an 
initial condition $P_0(n_0)$ evaluated along the trajectory $n_{(\tau)}$.\\
The time derivative of this trajectory entropy
\begin{eqnarray}
\dot{s}(\tau) = -\frac{\partial_{\tau} P_{\tau}(n)}{P_{\tau}(n)}\vert_{n_{(\tau)}}
- \sum_{j=1}^{N} \delta(\tau-\tau_j) \ln \frac{P_{\tau}(n_{j})}{P_{\tau}(n_{j-1})}
\;, \label{1Aaabb}
\end{eqnarray}
will be separated as
\begin{eqnarray}
\dot{s}(\tau) = \dot{s}_e(\tau) + \dot{s}_i(\tau) \;,
\label{1Aaabc}
\end{eqnarray}
where the trajectory entropy flux reads
\begin{eqnarray}
\dot{s}_e(\tau) \equiv - \sum_{j=1}^{N} \delta(\tau-\tau_j) 
\ln \frac{W_{\tau}(n_{j-1},n_{j})}{W_{\tau}(n_{j},n_{j-1})}
\label{1Aaabd}
\end{eqnarray}
and the trajectory entropy production reads
\begin{eqnarray}
\dot{s}_i(\tau) &\equiv& -\frac{\partial_t P_{\tau}(n)}{P_{\tau}(n)}\vert_{n_{(\tau)}} 
\label{1Aaabe} \\ &&- \sum_{j=1}^{N} \delta(\tau-\tau_j) \ln \frac{P_{\tau}(n_j) 
W_{\tau}(n_{j},n_{j-1})}{P_{\tau}(n_{j-1}) W_{\tau}(n_{j-1},n_{j})} \;. \nonumber
\end{eqnarray}
The ensemble average over the different trajectories is carried out by using the 
probability $P_{\tau}(n_{j-1}) W_{\tau}(n_{j-1},n_{j})$ that a transition occurs at 
time $\tau_j$ between $n_{j-1}$ and $n_{j}$. We get
\begin{eqnarray}
\dot{S}(\tau) &=& \langle \dot{s}(\tau) \rangle \\
\dot{S}_e(\tau) &=& \langle \dot{s}_e(\tau) \rangle \\
\dot{S}_i(\tau) &=& \langle \dot{s}_i(\tau) \rangle 
\label{1Aaabf}
\end{eqnarray}
The probability of a forward trajectory $n_{(\tau)}$ starting at time $0$
and ending at time $t$ is given by
\begin{eqnarray}
\mu_F[n_{(\tau)}] &=& P_0(n_0) \label{1Aaabg} \\
&&\hspace{-1.85cm} \left[ \prod_{j=1}^{N} 
\exp{\left( -\int_{\tau_{j-1}}^{\tau_{j}} d\tau' \sum_{m} 
W_{\tau'}(n_{j-1},m) \right)} W_{\tau_j}(n_{j-1},n_{j}) \right] 
\nonumber \\ &&\hspace{0.9cm}  
\exp{\left(-\int_{\tau_{N}}^{t} d\tau' 
\sum_{m} W_{\tau'}(n_{N},m)\right)} \nonumber \;.
\end{eqnarray}
The exponentials represent the probabilities to stay in a given 
state during the time interval between two successive jumps, and 
the transition rates evaluated at the jump times give the 
probability for the jumps to occur at this given times. \\
  
Defining the backward process as done for classical stochastic
dynamics \cite{Seifert} is not possible. 
In the classical case, it is sufficient after the forward process to 
revert the driving protocol $\tilde{\lambda}(\tau)=\lambda(t-\tau)$ 
and to ask for the probability of a backward trajectory (system taking 
the sequence of states of the forward trajectory but in the reversed 
order) to occur. 
The reversal of the driving protocol has the consequence of reversing 
the time dependence of the transition matrix 
$\tilde{W}_{\tau}(m,m')=W_{t-\tau}(m,m')$ so that the backward
process clearly correspond to a physical process.  
However, the time dependence of the quantum transition matrix $W_t(m,m')$ 
does not come exclusively from the external driving force and is different 
for different initial conditions of the subsystem $\hat{\sigma}(0)$.
Therefore, reversing the driving protocol does not simply
reverse the time dependence of the transition matrix. 
Nevertheless, in order to have a process corresponding to a reversal of 
the time dependence of the transition matrix, we will formally define an 
artificial backward process.
This definition will allow us to derive important 
fluctuation relations in next sections.\\       
Let us consider a new dynamics in the time interval 
$\tilde{\tau}=\lbrack 0,t \rbrack$ obeying 
\begin{eqnarray}
\dot{\tilde{P}}_{\tilde{\tau}}(m) = \sum_{\tilde{m}'} 
\tilde{W}_{\tilde{\tau}}(\tilde{m}',\tilde{m}) 
\tilde{P}_{\tilde{\tau}}(\tilde{m}') \;,
\label{1Aaabgb}
\end{eqnarray}
where the rates are related to the previous rates in the following way  
\begin{eqnarray}
\tilde{W}_{\tilde{\tau}}(\tilde{m}',\tilde{m})
=W_{t-\tilde{\tau}}(\tilde{m}',\tilde{m}) \;.
\label{1Aaabgc}
\end{eqnarray}
Let us call this dynamics with an arbitrary initial condition 
$\tilde{P}_0(\tilde{n}_0)$ the pseudo-backward dynamics.
This dynamic does not correspond in general to a quantum dynamics as 
(\ref{1Aaaao}).
We define now the following trajectory for this pseudo-backward dynamics 
\begin{eqnarray}
\tilde{n}_{(\tilde{\tau})} &=& 
\tilde{n}_0 \to \tilde{n}_1 \to \tilde{n}_2 \to \hdots \to \tilde{n}_N 
\nonumber \\ &=& n_N \to n_{N-1} \to n_{N-2} \to \hdots \to n_0 \;. 
\label{1Aaabgd}
\end{eqnarray}
where the jumps between $\tilde{n}_{j-1}$ and $\tilde{n}_j$ occur at 
time $\tilde{\tau}_j=t-\tau_{N-j+1}$ and where $\tilde{n}_j = n_{N-j}$.
Because this dynamics as well as the dynamics (\ref{1Aaaap}) both span
the same configuration space, summing over all trajectories of the 
pseudo-backward process is equivalent to summing over all the 
trajectories of the original process.
The trajectory (\ref{1Aaabgd}) is depicted on Fig. \ref{fig1}.
The forward probability of this trajectory is
evaluated in appendix \ref{pseudoback} and reads
\begin{eqnarray}
\tilde{\mu}_F[\tilde{n}_{(\tilde{\tau})}] &=& \tilde{P}_0(n_N) \label{1Aaabgh} \\
&&\hspace{-1.85cm} \bigg[ \prod_{j=1}^{N} 
\exp{\left( -\int_{\tau_{j-1}}^{\tau_{j}} d\tau' 
\sum_{m} W_{\tau'}(n_{j-1},m) \right)} 
\nonumber \\ &&\hspace{-1.2cm}  
W_{\tau_{j}}(n_{j},n_{j-1}) \bigg] 
\exp{\left(-\int_{\tau_N}^{t} d\tau' 
\sum_{m} W_{\tau'}(n_{N},m) \right)} \nonumber \;.
\end{eqnarray}
We now consider the ratio of the forward (\ref{1Aaabg}) and 
pseudo-backward (\ref{1Aaabgh}) probability. 
Noticing that the exponentials cancels, we find the fundamental 
result of the paper
\begin{eqnarray}
r(t)=\ln \frac{\mu_F[n_{(\tau)}]}{\tilde{\mu}_F[\tilde{n}_{(\tilde{\tau})}]} 
= \ln \frac{P_0(n_0)}{\tilde{P}_0(n_N)} - \Delta s_e(t) \;,
\label{1Aaabgi}
\end{eqnarray}
where the trajectory entropy flow is
\begin{eqnarray}
\Delta s_e(t) &\equiv& s_e(t)-s_e(0) = \int_{0}^{t} 
d\tau \dot{s}_e(\tau) \nonumber\\
&=& \sum_{j=1}^N \ln \frac{W_{\tau_{j}}(n_{j},n_{j-1})}
{W_{\tau_{j}}(n_{j-1},n_{j})} \;.
\label{1Aaabj}
\end{eqnarray}
In analogy with the classical results of Seifert \cite{Seifert}, 
we can now derive the various fluctuation theorems by specific choices of 
initial conditions for the pseudo-backward trajectories. 
By choosing $\tilde{P}_0(n_N)=P_t(n_N)$ and using the trajectory entropy 
\begin{eqnarray}
\Delta s(t) \equiv s(t)-s(0) =\ln \frac{P_0(n_0)}{P_t(n_N)} \;,
\label{1Aaabk}
\end{eqnarray}
Eq. (\ref{1Aaabgi}) becomes 
\begin{eqnarray}
r(t)=\ln \frac{\mu_F[n_{(\tau)}]}{\tilde{\mu}_F[\tilde{n}_{(\tilde{\tau})}]} 
= \Delta s(t) - \Delta s_e(t)  = \Delta s_i(t)
\label{1Aaabi}
\end{eqnarray}
where $\Delta s_i(t)=s_i(t)-s_i(0)$ is the trajectory entropy production. \\

Eq. (\ref{1Aaabgi}) has been first derived by Crooks \cite{Crooks1} for classical 
stochastic processes and later generalized by others \cite{Seifert,Maes3}.
We have shown that this relation may be extended to quantum systems.
The pseudo-backward trajectories are artificial.
In the classical case, because the time dependence of the rates 
is exclusively due to the external driving, the backward process 
has a physical meaning (e.g. \cite{Seifert}).
However, in the quantum case the time dependence of the rates is also due
to the quantum evolution of the density matrix itself, preventing us from 
associating in general a physical process to the backward dynamics.

\section{Quantum integral fluctuation theorem \label{integral}}

Summing over all possible trajectories of the pseudo-backward process is 
equivalent to summing over all possible trajectories of the original process 
$\sum_{\tilde{n}_{(\tilde{\tau})}}=\sum_{n_{(\tau)}}$. 
By averaging (\ref{1Aaabgi}) over all possible trajectories, we find 
\begin{eqnarray}
1&=&\sum_{\tilde{n}_{(\tilde{\tau})}} \tilde{\mu}_F[\tilde{n}_{(\tilde{\tau})}]
=\sum_{n_{(\tau)}} \tilde{\mu}_F[\tilde{n}_{(\tilde{\tau})}] \nonumber \\
&=&\sum_{n_{(\tau)}} \mu_F[n_{(\tau)}] e^{-r(t)} 
= \langle e^{-r(t)} \rangle \;.
\label{1Aaabl}
\end{eqnarray}
This integral fluctuation theorem \cite{Seifert} is valid for any choice of 
$P_0(n_0)$ and $\tilde{P}_0(n_N)$ in (\ref{1Aaabgi}).
Using the fact that $\langle e^x \rangle \geq e^{\langle x \rangle}$ this 
relation also means that in average the quantity $r(t)$ is always non-negative 
$\langle r(t) \rangle \geq 0$.
Choosing $\tilde{P}_0(n_N)=P_t(n_N)$ we have $r(t)=\Delta s_i(t)$ and we 
show again [see text below (\ref{1Aaaau})] that the ensemble averaged 
trajectory entropy production is always non-negative 
$\langle \Delta s_i(t) \rangle \geq 0$. 

\section{Quantum fluctuation theorem for steady state \label{steady}}

We consider a subsystem which is subjected to nonequilibrium constraints and
we assume that its dynamics can be described by a QME of the form (\ref{1Aaaaja}). 
When the subsystem is in a steady state, its density matrix does not evolve in time 
and the rates in equation (\ref{1Aaaap}) are time independent.  
An example of such system could be a two-level atom driven by a coherent single 
mode field on resonance (in the dipole approximation and in the rotating wave 
approximation) described by Bloch equations (see p154 of Ref. \cite{Breuer}).
In a steady state, the pseudo-backward process introduced in section 
\ref{trajectory} would correspond to the real physical backward process 
\begin{eqnarray}
\tilde{\mu}_F[\tilde{n}_{(\tilde{\tau})}]=\mu_B[n_{(\tau)}] \;.
\label{1Aaabma}
\end{eqnarray}
By definition, we have 
\begin{eqnarray}
p_F(R(t)) &=& \langle \delta(R(t) - r_F(t) ) \rangle_F \nonumber \\
&=&\sum_{n_{(\tau)}} \mu_F[n_{(\tau)}] \delta(R(t) - r_F(t)) 
\nonumber 
\label{1Aaabmb}
\end{eqnarray}
Using (\ref{1Aaabgi}), we can write
\begin{eqnarray}
p_F(R(t)) &=& \sum_{n_{(\tau)}} \mu_B[n_{(\tau)}] e^{r_F(t)} 
\delta(R(t) - r_F(t)) \nonumber \\
&=& \sum_{n_{(\tau)}} \mu_B[n_{(\tau)}] e^{R(t)} 
\delta(R(t) - r_F(t)) \nonumber \\
&=& \langle \delta(R(t)+r_B(t)) \rangle_B \; e^{R(t)} \nonumber \\ 
&=& p_B(-R(t)) \; e^{R(t)} \;, 
\label{1Aaabmc}
\end{eqnarray}
where to go from the second line to the third one, we used 
$r_F(t)=-r_B(t)$ which comes from (\ref{1Aaabma}) with (\ref{1Aaabgi}). 
When $\tilde{P}_0(n_N)=P_t(n_N)$ and therefore Eq. (\ref{1Aaabi}) holds, 
Eq. (\ref{1Aaabmc}) becomes a fluctuation theorem for the entropy production 
\begin{eqnarray}
p_F(\Delta S_i(t))=p_B(-\Delta S_i(t)) \; e^{\Delta S_i(t)}
\label{1Aaabmd}
\end{eqnarray}
This relation shows that at steady state, the ratio of the probability 
to observe a given entropy production during a froward process and the probability 
to observe the same entropy production with a minus sign during the backward 
process is given by the exponential of the entropy production. 
This is the most familiar form of the fluctuation theorem.
In the infinite time limit, if the subsystem as a finite number of levels 
(this condition is usually implicitly assumed in QME theory) $\Delta S(t)$
will be bounded and $\Delta S_i(t) \stackrel{t \to \infty}{=} \Delta S_e(t)$
so that (\ref{1Aaabmd}) also becomes a fluctuation theorem for the entropy flow
and therefore also for the heat.
For completeness, we give in appendix \ref{MonnaiTh} a different derivation 
of a fluctuation theorem similar to (\ref{1Aaabmd}) and which is not restricted 
to steady states. 

\section{Heat and work for quantum trajectories \label{trajQW}}

If we use the relation (\ref{1Aaaay}) together with the definition of 
the trajectory entropy flow (\ref{1Aaabj}), we find that the heat 
associated with a single trajectory is given by
\begin{eqnarray}
\tilde{q}_S(t) &\equiv& \beta^{-1} \Delta s_e(t) \label{1Aaabib} \\
&=& \sum_{j=1}^{N} \left( \bra{n_{j}} \hat{H}_S(\tau_{j}) \ket{n_{j}} 
- \bra{n_{j-1}} \hat{H}_S(\tau_{j}) \ket{n_{j-1}} \right) \;.
\nonumber
\end{eqnarray}
The interpretation of this result is that the heat flowing to the 
environment results from transitions between the subsystem states $n_j$.\\ 
The energy associated with a trajectory is a state function and only 
depends on the initial and final state of the trajectory  
\begin{eqnarray}
\Delta e_S(t)&=&\bra{n_N} \hat{H}_S(t) \ket{n_N} 
- \bra{n_0} \hat{H}_S(0) \ket{n_0} 
\nonumber\\
&=& \sum_{j=1}^{N} \left( \bra{n_{j}} \hat{H}_S(\tau_{j}) \ket{n_{j}} 
- \bra{n_{j-1}} \hat{H}_S(\tau_{j}) \ket{n_{j-1}} \right) \nonumber\\
&=& \tilde{w}_S(t) + \tilde{q}_S(t) \;.
\label{1Aaabic}
\end{eqnarray}
The work is therefore given by
\begin{eqnarray}
\tilde{w}_S(t)&=& \Delta e_S(t) - \tilde{q}_S(t) \label{1Aaabicb}\\
&&\hspace{-1.3cm} =
\sum_{j=1}^{N} \left( \bra{n_{j-1}} \hat{H}_S(\tau_{j}) \ket{n_{j-1}}
- \bra{n_{j-1}} \hat{H}_S(\tau_{j-1}) \ket{n_{j-1}} \right) \nonumber \;.
\end{eqnarray}
The work thus results from the
time evolution of the Hamiltonian (due to the driving force) along
the states $n_j$ of the subsystem between the transitions.
It is interesting to make the parallel between our description of 
heat and work in the $\{\ket{m_t}\}$ basis set and the adiabatic basis 
description of Ref. \cite{Mukamel}. In the latter the work comes from
the evolution along the adiabatic states and the heat comes
from the transitions between the adiabatic state. 
This can be understood by comparing (\ref{Apaai}) and (\ref{Apaaj})
with (\ref{Apaam}) and (\ref{Apaan}).

\section{The quantum Jarzynski relation \label{Jarzynski}}

We assume that the subsystem is initially at equilibrium with respect
to the Hamiltonian $\hat{H}_S(0)=\hat{H}_S(\lambda(0))$ and is therefore 
described by a canonical distribution. 
The system is then driven out of equilibrium by turning the driving 
force from $\lambda(0)$ to $\lambda(t')$ at time $t'$.
After $t'$ the driving force stop evolving.
On long time scales after $t'$, say $t$ ($t \gg t'$), the system is again 
at equilibrium in a canonical distribution but now with respect to 
$\hat{H}_S(t)=\hat{H}_S(\lambda(t))$.\\
We choose
\begin{eqnarray}
P_0(n_0) &=& \frac{e^{-\beta \bra{n_0} \hat{H}_S(0) \ket{n_0}}}{Z_0} \nonumber \\
P_{t}(n_N) &=& \frac{e^{-\beta \bra{n_N} \hat{H}_S(t) \ket{n_N}}}{Z_{t}} \;,
\label{1Aaabpb}
\end{eqnarray}
where $Z_0=\sum_n \exp{(-\beta \bra{n_0} \hat{H}_S(0) \ket{n_0})}$
and $Z_t=\sum_n \exp{(-\beta \bra{n_N} \hat{H}_S(t) \ket{n_N})}$.
Notice that $\{\ket{n_0}\}$ [$\{\ket{n_t}\}$] is now the 
eigenbasis of $\hat{H}_S(0)$ [$\hat{H}_S(t)$].\\
The free energy difference between the initial
and the final state is given by  
\begin{eqnarray}
\Delta F(t)= F(t)-F(0) = - \beta^{-1} \ln \frac{Z_t}{Z_0} \;.
\label{1Aaabpa}
\end{eqnarray}
Using Eq. (\ref{1Aaabib}) which defines the heat of a single subsystem 
trajectory, we can write Eq. (\ref{1Aaabi}) as 
\begin{eqnarray}
\Delta s_i(t) &=& - \ln P_t(n_N) + \ln P_0(n_0) - \beta  \tilde{q}_S(t) \;.
\label{1Aaabo}
\end{eqnarray}
Using now (\ref{1Aaabpb}), (\ref{1Aaabpa}), (\ref{1Aaabic}) 
and (\ref{1Aaabicb}), we can rewrite (\ref{1Aaabo}) as
\begin{eqnarray}
\Delta s_i(t) &=& - \beta \Delta F(t) +\beta \tilde{w}_S(t) \;.
\label{1Aaabq}
\end{eqnarray}
Finally, by inserting Eq. (\ref{1Aaabq}) in the integral fluctuation 
theorem (\ref{1Aaabl}) where $r(t)=\Delta s_i(t)$, we find the quantum 
Jarzynski relation
\begin{eqnarray}
e^{- \beta \Delta F(t)} = \langle e^{- \beta \tilde{w}_S(t)} \rangle \;.
\label{1Aaabt}
\end{eqnarray}

\section{Conclusions \label{conclusion}}

We have presented a unified derivation of a quantum integral 
fluctuation theorem, a quantum steady state fluctuation theorem, 
and the quantum Jarzynski relation, for a driven subsystem
interacting with its environment and described by a QME.
This generalizes earlier results obtained for quantum systems.
By recasting the solution of the QME in a BDME form with time dependent 
rate for the eigenvalues of the subsystem density matrix, we naturally 
define quantum trajectories and their associated entropy, heat and 
work and study their fluctuation properties. 
The connection between the trajectory quantities which naturally 
enter our formulation and measurable quantum trajectory quantities 
is still an open issue. 
Deriving quantum fluctuation relations without having to assume QME, 
which do not correctly account for strong subsystem-environment 
entanglement, is an exiting perspective. 


\begin{acknowledgments}
The support of the National Science Foundation (Grant No. CHE-0446555),
NIRT (Grant No. EEC 0303389), the National Institute of Health 
(Grant No. GM59230-05) is gratefully acknowledged.
M. E. is also supported by the FNRS Belgium 
(collaborateur scientifique).\\
\end{acknowledgments}

\appendix

\section{Basis dependence of heat and work \label{appQW}}

We consider a system with a time dependent Hamiltonian 
$\hat{H}(t)$ in the Schrodinger picture described 
by the density matrix $\hat{\rho}(t)$.
The evolution equation of $\hat{\rho}(t)$ is not necessarily unitary.\\
The energy of the system is given by
\begin{eqnarray}
\mean{\hat{H}} \equiv \trace \hat{H}(t) \hat{\rho}(t) 
= \sum_{aa'} \bra{a_t} \hat{H}(t) \ket{a_t'} \bra{a_t'} \hat{\rho}(t) \ket{a_t} \;,
\label{Apaaa}
\end{eqnarray}
where $\{ \ket{a_t} \}$ is an arbitrary time dependent basis set.
The energy changes of the system can be written as
\begin{eqnarray}
\Delta E(t) &\equiv& \int_{0}^{t} d\tau \frac{d \mean{\hat{H}(\tau)}}{d\tau} \\
&=& W(t) + Q(t) \label{Apaab}
= \tilde{W}(t) + \tilde{Q}(t) \nonumber
\end{eqnarray}
where the heat and the work are given by
\begin{eqnarray}
\dot{Q}(t) &\equiv& \trace \hat{H}(t) \dot{\hat{\rho}}(t) \label{Apaac}\\
\dot{W}(t) &\equiv& \trace \dot{\hat{H}}(t) \hat{\rho}(t) \label{Apaad}\;,
\end{eqnarray}
in a time independent basis and by
\begin{eqnarray}
\dot{\tilde{Q}}(t) &\equiv& \sum_{aa'} \bra{a_t} \hat{H}(t) \ket{a_t'} 
\frac{d}{dt} \bigg( \bra{a_t'} \hat{\rho}(t) \ket{a_t} \bigg) \label{Apaae}\\
\dot{\tilde{W}}(t) &\equiv& \sum_{aa'} 
\frac{d}{dt} \bigg( \bra{a_t} \hat{H}(t) \ket{a_t'} \bigg) 
\bra{a_t'} \hat{\rho}(t) \ket{a_t} \;, \label{Apaaf}
\end{eqnarray}
in a time dependent basis.\\
How does $Q(t)$ [$W(t)$] relates to $\tilde{Q}(t)$ [$\tilde{W}(t)$]?
We find
\begin{eqnarray}
\dot{\tilde{Q}}(t) 
&=& \dot{Q}(t) - \dot{A}(t) \label{Apaag}\\
\dot{\tilde{W}}(t) 
&=& \dot{W}(t) + \dot{A}(t) \;,\nonumber
\end{eqnarray}
where
\begin{eqnarray}
\dot{A}(t)&=& - \sum_{aa'} \bra{a_t} \hat{H}(t) \ket{a_t'} \bigg( \bra{\dot{a}_t'} 
\hat{\rho}(t) \ket{a_t} + \bra{a_t'} \hat{\rho}(t) \ket{\dot{a}_t} \bigg) \nonumber\\
&=& \sum_{aa'} \bigg( \bra{\dot{a}_t} \hat{H}(t) \ket{a_t'} 
+ \bra{a_t} \hat{H}(t) \ket{\dot{a}_t'} \bigg) \bra{a_t'} \hat{\rho}(t) \ket{a_t} \;.
\nonumber \\ \label{Apaah}
\end{eqnarray}
We have used the fact that $\braket{a_t}{\dot{a}_t'}=-\braket{\dot{a}_t}{a_t'}$ 
which come from $\frac{d}{dt} (\braket{a_t}{a_t'})=0$.\\
If we consider the time dependent basis set which diagonalizes the instantaneous 
density density matrix $\{ \ket{a_t} \} = \{ \ket{m_t} \}$, where 
$\bra{m_t} \hat{\rho}(t) \ket{m_t'}=P_t(m) \delta_{mm'}$, we have
\begin{eqnarray}
\dot{\tilde{Q}}(t) 
&=& \sum_{m} \bra{m_t} \hat{H}(t) \ket{m_t} \dot{P}_t(m) \label{Apaai}\\
&=& \dot{Q}(t) - \dot{A}(t) \nonumber\\
\dot{\tilde{W}}(t)
&=& \sum_{m} \frac{d}{dt} 
\bigg( \bra{m_t} \hat{H}(t) \ket{m_t} \bigg) P_t(m) \label{Apaaj}\\
&=& \dot{W}(t) + \dot{A}(t) \;,\nonumber
\end{eqnarray}
where
\begin{eqnarray}
\dot{A}(t) 
&=& \sum_{mm'} \bra{m_t} \hat{H}(t) \ket{m_t'} 
\braket{m_t'}{\dot{m}_t} \bigg( P_t(m) - P_t(m') \bigg) \nonumber \\
&=& \sum_{m} \bigg( \bra{\dot{m}_t} \hat{H}(t) \ket{m_t} 
+ \bra{m_t} \hat{H}(t) \ket{\dot{m}_t} \bigg) P_t(m) \nonumber \;.\\
\label{Apaal}
\end{eqnarray}
If we consider the time dependent basis diagonalizing the instantaneous 
Hamiltonian (adiabatic basis) $\{ \ket{a_t} \} = \{ \ket{i_t} \}$, where 
$\bra{i_t} \hat{H}(t) \ket{i_t'}=\epsilon_i(t) \delta_{ii'}$, we have 
\begin{eqnarray}
\dot{\tilde{Q}}'(t) 
&=& \sum_{i} \epsilon_i(t) \frac{d}{dt} 
\bigg( \bra{i_t} \hat{\rho}(t) \ket{i_t} \bigg) \label{Apaam}\\
&=& \dot{Q}(t) - \dot{A}'(t) \nonumber\\
\dot{\tilde{W}}'(t) 
&=& \sum_{i} \dot{\epsilon}_i(t) \bra{i_t} \hat{\rho}(t) \ket{i_t} \label{Apaan}\\ 
&=& \dot{W}(t) + \dot{A}'(t) \;,\nonumber
\end{eqnarray}
where
\begin{eqnarray}
\dot{A}'(t) &=& - \sum_{i} \epsilon_i(t) \bigg( \bra{\dot{i}_t} \hat{\rho}(t) 
\ket{i_t} + \bra{i_t} \hat{\rho}(t) \ket{\dot{i}_t} \bigg) \label{Apaao}\\
&=& \sum_{ii'} \bigg( \epsilon_i(t) - \epsilon_{i'}(t) \bigg) 
\braket{i_t}{\dot{i}_t'} \bra{i_t'} \hat{\rho}(t) \ket{i_t} \nonumber \;.
\end{eqnarray}
It is interesting to notice the similarity between the two basis 
$\{ \ket{m_t} \}$ and $\{ \ket{i_t} \}$. 
In both cases, the heat results from changes in the population of the 
states (and therefore from transitions between states) [see (\ref{Apaai}) 
and (\ref{Apaam})] and the work from the evolution of the
Hamiltonian along the states [see (\ref{Apaaj}) and (\ref{Apaan})].   
Using (\ref{Apaal}) with (\ref{Apaao}) one gets
\begin{eqnarray}
\dot{A}(t) - \dot{A}'(t) &=& \dot{\tilde{W}}(t) - \dot{\tilde{W}}'(t) \label{Apaaob} \\
&=& \dot{\tilde{Q}}'(t) - \dot{\tilde{Q}}(t) \nonumber \\
&=& \sum_{i,m} P_t(m) \epsilon_i(t) \frac{d}{dt} 
\bigg( \vert \braket{i_t}{m_t} \vert^2 \bigg) \nonumber \;.
\end{eqnarray}
Let us assume now that the density matrix of the system obeys the von 
Neumann equation 
\begin{eqnarray}
\dot{\hat{\rho}}(t)=-\i[\hat{H}(t),\hat{\rho}(t)]
=\hat{\hat{{\cal L}}}(t) \hat{\rho}(t) \;, \label{Apaap}
\end{eqnarray}
whose solution reads
\begin{eqnarray}
\hat{\rho}(t)&=& \hat{\hat{{\cal U}}}(t) \hat{\rho}(0) 
= \hat{U}(t) \hat{\rho}(0) \hat{U}^{\dagger}(t) \;,\label{Apaaq}
\end{eqnarray}
where
\begin{eqnarray}
\hat{\hat{{\cal U}}}(t) &=& \exp_{+}{\{\int_{0}^{t} d\tau 
\hat{\hat{{\cal L}}}(\tau)\}} \label{Apaar}\\
\hat{U}(t) &=& \exp_{+}{\{-\i \int_{0}^{t} d\tau \hat{H}(\tau)\}} \;. \nonumber
\nonumber
\end{eqnarray}
The evolution operator [superoperator] $\hat{U}_T(t)$ [$\hat{\hat{{\cal U}}}_T(t)$] 
is unitary.
In this case, the expression in the basis $\{ \ket{m_t} \}$ simplify to
\begin{eqnarray}
\tilde{Q}(t) &=& Q(t) = 0 \label{Apaas}\\
\tilde{W}(t) &=& W(t) = \Delta E (t) \;. \label{Apaat}
\end{eqnarray}
This is due to the fact that 
\begin{eqnarray}
P_0(m) &=& \bra{m_0} \hat{\rho}(0) \ket{m_0} \label{Apaau}\\ 
&=& \bra{m_0} \hat{U}^{\dagger}(t) \hat{U}(t) \hat{\rho}(0) 
\hat{U}^{\dagger}(t) \hat{U}(t) \ket{m_0} \nonumber \\
&=& \bra{m_t} \hat{\rho}(t) \ket{m_t} = P_t(m) \;.\nonumber
\end{eqnarray}
This means that no heat is produced by the driving force for a unitary evolution. 
This is reasonable since there is no environment. The only way in which the energy
of the system may increase is via the work done on the system. 
Notice that in the adiabatic basis both $\tilde{W}'(t)$ and
$\tilde{Q}'(t)$ are finite for a unitary evolution.

\section{Probability of the pseudo-backward trajectory  \label{pseudoback}}

The probability of a pseudo-backward trajectory $\tilde{n}_{(\tilde{\tau})}$ reads
\begin{eqnarray}
\tilde{\mu}_F[\tilde{n}_{(\tilde{\tau})}] &=& \tilde{P}_0(\tilde{n}_0) \label{1Aaabge} \\
&&\hspace{-1.85cm} \bigg[ \prod_{j=1}^{N} 
\exp{\left( -\int_{\tilde{\tau}_{j-1}}^{\tilde{\tau}_{j}} d\tilde{\tau}' 
\sum_{\tilde{m}} \tilde{W}_{\tilde{\tau}'}(\tilde{n}_{j-1},\tilde{m}) \right)} 
\tilde{W}_{\tilde{\tau}_j}(\tilde{n}_{j-1},\tilde{n}_{j}) \bigg] 
\nonumber \\ &&\hspace{0.9cm}  
\exp{\left(-\int_{\tilde{\tau}_{N}}^{t} d\tilde{\tau}' 
\sum_{\tilde{m}} \tilde{W}_{\tilde{\tau}'}(\tilde{n}_{N},\tilde{m})\right)} 
\nonumber \;,
\end{eqnarray}
where $\tilde{\tau}_0=0$ and $\tilde{\tau}_{N+1}=t$. 
Using (\ref{1Aaabgc}), (\ref{1Aaabgd}) and $\tilde{\tau}_j=t-\tau_{N-j+1}$, 
we can rewrite this probability as
\begin{eqnarray}
\tilde{\mu}_F[\tilde{n}_{(\tilde{\tau})}] &=& \tilde{P}_0(n_N) \label{1Aaabgf} \\
&&\hspace{-1.85cm} \bigg[ \prod_{j=1}^{N} 
\exp{\left( -\int_{t-\tau_{N-j+2}}^{t-\tau_{N-j+1}} d\tilde{\tau}' 
\sum_{m} W_{t-\tilde{\tau}'}(n_{N-j+1},m) \right)} 
\nonumber \\ &&\hspace{2.5cm}  
W_{\tau_{N-j+1}}(n_{N-j+1},n_{N-j}) \bigg] 
\nonumber \\ &&\hspace{0cm}  
\exp{\left(-\int_{t-\tau_1}^{t} d\tilde{\tau}' 
\sum_{m} W_{t-\tilde{\tau}'}(n_{0},m)\right)} 
\nonumber \;.
\end{eqnarray}
Using the change of variable $\tau=t-\tilde{\tau}$, we get
\begin{eqnarray}
\tilde{\mu}_F[\tilde{n}_{(\tilde{\tau})}] &=& \tilde{P}_0(n_N) 
\exp{\left(-\int_{0}^{\tau_1} d\tau' 
\sum_{m} W_{\tau'}(n_{0},m) \right)} \nonumber \\
&&\hspace{-1.8cm} \bigg[ \prod_{j=1}^{N} 
\exp{\left( -\int_{\tau_{N-j+1}}^{\tau_{N-j+2}} d\tau' 
\sum_{m} W_{\tau'}(n_{N-j+1},m) \right)}
\nonumber \\ &&\hspace{1.6cm}  
W_{\tau_{N-j+1}}(n_{N-j+1},n_{N-j}) \bigg] \label{1Aaabgg}\;.
\end{eqnarray}
With help of $j=N-j_{old}+2$, (\ref{1Aaabgg}) finally becomes (\ref{1Aaabgh}).

\section{Quantum fluctuation theorem for uncorrelated 
subsystem and bath \label{MonnaiTh}}

We derive a general quantum fluctuation theorem (not restricted to steady states) 
for a driven quantum subsystem in contact with its environment. 
The derivation is similar to the derivation of Monnai in \cite{Monnai}
and is given for completeness. \\

We assume weak coupling between the subsystem and the environment
and that the environment is infinitely large so that 
at all times the density matrix of the total system 
(subsystem plus environment) can be written as  
\begin{eqnarray}
\hat{\rho}(t) = \hat{\sigma}(t) \hat{\rho}_B^{eq} \;,
\label{1Aaaca}
\end{eqnarray}
where $\hat{\rho}_B^{eq}=e^{-\beta \hat{H}_B}/Z_B$ is the time independent 
equilibrium reduced density matrix of the environment and 
$\hat{\sigma}(t)$ the time dependent reduced density matrix of the subsystem.
Assuming the form (\ref{1Aaaca}) is not very different from assuming
that the subsystem density matrix obeys a QME since most of the QME 
derivation implicitly assume an invariant environment density matrix
(e.g. the Born approximation \cite{Breuer}). \\
Let us define the basis $\{\ket{m_t b}\}$, where $\{\ket{m_t}\}$ diagonalize 
the subsystem density matrix at time $t$ and where $\{\ket{b}\}$ 
diagonalize the time independent environment Hamiltonian.
The probability to go from $\ket{m_0 b}$ at time $0$ to $\ket{m_t b'}$ 
at time $t$ is given by
\begin{eqnarray}
\mu_F[\ket{m_0 b} \to \ket{m_t b'}] &=& \label{1Aaacb}\\
&&\hspace{-1.5cm} \bra{m_0 b} \hat{\sigma}(0) \hat{\rho}_B^{eq} \ket{m_0 b}
\vert \bra{m_t b'} \hat{U}(t) \ket{m_0 b} \vert^2 \;, \nonumber
\end{eqnarray}
where $\hat{U}(t)$ is the unitary evolution operator of the total system. 
The probability of the backward process to go from $\ket{m_t b'}$ 
at time $t$ to $\ket{m_0 b}$ at time $0$ by the time reversed evolution 
\cite{Maes2,Maes4,Monnai} is given by 
\begin{eqnarray}
\mu_B[\ket{m_t b'} \to \ket{m_0 b}] &=& \label{1Aaacc}\\
&&\hspace{-1.5cm} \bra{m_t b'} \hat{\sigma}(t) \hat{\rho}_B^{eq} \ket{m_t b'}
\vert \bra{m_t b'} \hat{U}(t) \ket{m_0 b} \vert^2 \;. \nonumber
\end{eqnarray}
We therefore have that
\begin{eqnarray}
\frac{\mu_F[\ket{m_0 b} \to \ket{m_t b'}]}
{\mu_B[\ket{m_t b'} \to \ket{m_0 b}]} =
\frac{P_0(m_0)}{P_t(m_t)} e^{- \beta Q_{bb'}} \;,
\label{1Aaace}
\end{eqnarray}
where $P_0(m_0)=\bra{m_0} \hat{\sigma}(0) \ket{m_0}$, 
$P_t(m_t)=\bra{m_t} \hat{\sigma}(t) \ket{m_t}$ and $Q_{bb'}= E_b-E_{b'}$.\\
The entropy of a state $m_t$ is defined as 
\begin{eqnarray}
s(t)=-\ln P_t(m_t) \;.
\label{1Aaacf}
\end{eqnarray}
This definition makes sense because $\{ \ket{m_t} \}$ diagonalizes $\hat{\sigma}(t)$,
so that by averaging over the different states we recover the von Neumann entropy.
The entropy difference between the initial and the final state of the subsystem 
starting at time $0$ in $\ket{m_0}$ and ending at time $t$ in $\ket{m_t}$ 
is given by
\begin{eqnarray}
\Delta s(s_0,s_t;t)= \ln \frac{P_0(m_0)}{P_t(m_t)}
\label{1Aaacg}
\end{eqnarray}
The entropy production of this same process is given by
\begin{eqnarray}
\Delta s_i(m_0,m_t,b,b';t) = 
\ln \frac{P_0(m_0)}{P_t(m_t)} - \frac{Q_{bb'}}{T} \;,
\label{1Aaach}
\end{eqnarray}
because one assumes that the entropy flow difference is given by
\begin{eqnarray}
\Delta s_e(b,b';t) = \frac{Q_{bb'}}{T} \;.
\label{1Aaaci}
\end{eqnarray}
Using (\ref{1Aaacg}), (\ref{1Aaach}) and (\ref{1Aaaci}), 
Eq. (\ref{1Aaace}) becomes 
\begin{eqnarray}
\ln \frac{\mu_F[\ket{m_0 b} \to \ket{m_t b'}]}
{\mu_B[\ket{m_t b'} \to \ket{m_0 b}]} &=& \Delta s_i(m_0,m_t,b,b';t) 
\label{1Aaacj} \\ 
&=& \Delta s(s_0,s_t;t) - \Delta s_e(b,b';t) \;. \nonumber
\end{eqnarray}
This result is the analog of our fundamental relation (\ref{1Aaabi}).
By averaging the probabilities over all possible initial and final states, 
we get the general fluctuation theorem 
\begin{eqnarray}
p(\Delta S_i(t)) &=& 
\sum_{m_0,m_t,b,b'} \mu_F[\ket{m_0 b} \to \ket{m_t b'}] \nonumber \\
&&\hspace{1.1cm} \delta(\Delta S_i(t)-\Delta s_i(m_0,m_t,b,b';t)) \nonumber \\
&&\hspace{-1.5cm}= \sum_{m_0,m_t,b,b'} 
\mu_B[\ket{m_0 b} \to \ket{m_t b'}] \nonumber \\
&&\hspace{-1cm} e^{\Delta s_i(m_0,m_t,b,b';t)} 
\delta(\Delta S_i(t)-\Delta s_i(m_0,m_t,b,b';t)) \nonumber \\
&&\hspace{-1.5cm}= p(-\Delta S_i(t)) e^{\Delta S_i(t)} \;.
\label{1Aaack}
\end{eqnarray}
This result agrees with (\ref{1Aaabmd}) and is not restricted to steady states.
This approach is based on the time reversal invariance of the evolution of 
the total system and does not provide a trajectory picture.  
We note that in the total system space, the heat (or the entropy flow) going 
from the subsystem to the environment only depends on the end points and not 
on the path itself. 
If one derive a Jarzynski relation from this result as in \cite{Monnai}, the 
work is also path independent.
When considering the reduced dynamics of the system alone, as done in this paper, 
these quantities become path dependent and a trajectory picture is provided.



\begin{thebibliography} {1}

\bibitem{Jarzynski1}
C. Jarzynski, Phys. Rev. Lett. {\bf 78}, 2690-2693 (1997);
C. Jarzynski, Phys. Rev. E {\bf 56}, 5018-5035 (1997).

\bibitem{Crooks0}
G. E. Crooks, J. Stat. Phys. {\bf 90}, 1481 (1998).

\bibitem{JarzynskiReply}
C. Jarzynski, J. Stat. Mech. (2004) P09005.

\bibitem{Cohen}
E G D Cohen and D. Mauzerall, J. Stat. Mech. (2004) P07006.

\bibitem{Maes3}
C. Maes, S\'eminaire Poincar\'e {\bf 2}, 29 (2003).

\bibitem{Evans1}
D. J. Evans, E. G. D. Cohen, and G. P. Morriss, 
Phys. Rev. Lett. {\bf 71}, 2401 (1993).

\bibitem{Gallavotti}
G. Gallavotti and E. G. D. Cohen, Phys. Rev. Lett. {\bf 74}, 2694 (1995);
G. Gallavotti and E. G. D. Cohen, J. Stat. Phys. {\bf 80}, 931 (1995).

\bibitem{Evans2}
D. J. Evans and D. J. Searles, Phys. Rev. E 50, 1645 (1994);
D. J. Evans and D. J. Searles, Phys. Rev. E 52, 5839 (1995);
D. J. Evans and D. J. Searles, Phys. Rev. E 53, 5808 (1996).

\bibitem{Bochkov}
G. N. Bochkov and Yu. E. Kuzovlev, Zh. Eksp. Teor. Fiz. {\bf 72}, 238 (1977)
[Sov. Phys. JETP {\bf 45}, 125 (1977)];
G. N. Bochkov and Yu. E. Kuzovlev, Zh. Eksp. Teor. Fiz. {\bf 76}, 1071 (1979)
[Sov. Phys. JETP {\bf 49}, 543 (1979)];
G. N. Bochkov and Yu. E. Kuzovlev, Physica A {\bf 106}, 443 (1981);
G. N. Bochkov and Yu. E. Kuzovlev, Physica A {\bf 106}, 480 (1981).

\bibitem{Stratonovich}
R. L. Stratonovich, {\it Nonlinear nonequilibrium thermodynamics II}
(Springer, Berlin, 1994).

\bibitem{JarsynskiFT}
C. Jarsynski, J. Stat. Phys. {\bf 98}, 77 (2000).

\bibitem{Kurchan1}
J. Kurchan, J. Phys. A {\bf 31}, 3719 (1998).

\bibitem{Lebowitz}
J. L. Lebowitz and H. Spohn, J. Stat. Phys. {\bf 95}, 333 (1999).

\bibitem{Searles}
D. J. Searles and D. J. Evans, Phys. Rev. E {\bf 60}, 159 (1999).

\bibitem{Gaspard1}
P. Gaspard, J. Chem. Phys. {\bf 120}, 19 (2004);
D. Andrieux and P. Gaspard, J. Chem. Phys. {\bf 121}, 6167 (2004). 

\bibitem{Gaspard2}
P. Gaspard, J. Stat. Phys. {\bf 117}, 599 (2004);
Pierre Gaspard, Lectures notes for the International Summer School 
\textit{Fundamental Problems in statistical Physics XI} 
(Leuven, Belgium, September 4-17, 2005). 

\bibitem{Crooks1}
G. E. Crooks, Phys. Rev. E {\bf 60}, 2721 (1999). 

\bibitem{Crooks2}
G. E. Crooks, Phys. Rev. E {\bf 61}, 2361 (2000).

\bibitem{Seifert}
U. Seifert, Phys. Rev. Lett. {\bf 95}, 040602 (2005).

\bibitem{Yukawa}
S. Yukawa, J. Phys. Soc. Jpn. {\bf 69}, 2367 (2000).

\bibitem{HTasaki}
H. Tasaki, cond-mat/0009244 (2000).

\bibitem{Mukamel}
S. Mukamel, Phys. Rev. Lett. {\bf 90}, 170604 (2003). 

\bibitem{Maes}
W. De Roeck and C. Maes, Phys. Rev. E {\bf 69}, 026115 (2004).

\bibitem{Kurchan}
J. Kurchan, cond-mat/0007360 (2000).
 
\bibitem{STasaki}
T. Monnai and S. Tasaki, cond-mat/0308337 (2003).

\bibitem{Maes1}
W. De Roeck and C. Maes, cond-mat/0406004 (2004).

\bibitem{Monnai}
T. Monnai, Phys. Rev. E {\bf 72}, 027102 (2005). 

\bibitem{JarzynskiQ}
C. Jarzynski and D. K. Wojcik, Phys. Rev. Lett. {\bf 92}, 230602 (2004).

\bibitem{Allahverdyan}
A. E. Allahverdyan and Th. M. Nieuwenhuizen, Phys. Rev. E {\bf 71}, 066102 (2005).

\bibitem{Maes2}
I. Callens, W. De Roeck, T. Jacobs, C. Maes and K. Netocny,
Physica D {\bf 187}, 383 (2004).

\bibitem{Haake} 
F. Haake, {\it Statistical Treatment of Open Systems}, Springer Tracts
in Modern Physics, Vol. 66 (1973).

\bibitem{Spohn}
H. Spohn, Rev. Mod. Phys. {\bf 53}, 569 (1980).

\bibitem{Gardiner} 
C.W. Gardiner and P. Zoller, {\it Quantum Noise} (Springer, Berlin, 2000).

\bibitem{Breuer} 
H.-P. Breuer and F. Petruccione {\it The Theory of Open Quantum Systems}
(Oxford University Press, Oxford, 2002).

\bibitem{Comment}
The trace invariance used in Eq. (\ref{2Aaaadb}) could be a delicate matter
for systems with continuous spectrum where the trace may diverge. 
This can be always formally addressed by using a dense quasicontinuum.
Physically, Eq. (\ref{2Aaaadb}) implies energy conservation.

\bibitem{MukamelB}
S. Mukamel, Principles of nonlinear optical spectroscopy
(Oxford University Press, New York, 1995).

\bibitem{Prigogine}
D. Kondepudi and I. Prigogine {\it Modern thermodynamics} 
(Wiley, Chichester, 1998).

\bibitem{GrootMazur}
S. R. de Groot and P. Mazur, {\it Non-equilibrium thermodynamics} 
(Dover, New York, 1984). 

\bibitem{Kampen}
N. G. van Kampen, {\it Stochastic Processes in Physics and Chemistry}, 
2nd ed. (North-Holland, Amsterdam, 1997).

\bibitem{Maes4}
T. Yacobs and C. Maes, quant-ph/0508041 (2005).

\end{thebibliography}
\end{document}